\documentstyle[aps,prl,psfig,epsfig]{revtex}
\newcommand{\be}{\begin{equation}}
\newcommand{\ee}{\end{equation}}
\newcommand{\bea}{\begin{eqnarray}}
\newcommand{\eea}{\end{eqnarray}}

\tightenlines
\begin{document}
\draft

\title{
On the statistical mechanics of prion diseases}

\author{A. Slepoy, R.R.P. Singh, F. P\'{a}zm\'{a}ndi\cite{ferencaddresses}, R.N. Kulkarni and D.L. Cox }
\address{
Department of Physics, University of California, Davis, CA 95616}
\twocolumn[\hsize\textwidth\columnwidth\hsize\csname
@twocolumnfalse\endcsname

\date{\today}

\maketitle

\begin{abstract}
We simulate a two-dimensional, lattice based,  protein-level
statistical mechanical model for prion diseases (e.g., Mad Cow disease)
with concommitant prion protein misfolding and aggregation.
Our simulations lead us to the hypothesis that
the observed broad incubation time distribution in
epidemiological data reflect fluctuation dominated growth seeded by a few
nanometer scale aggregates, while much narrower
incubation time distributions for innoculated lab
animals arise from statistical
self averaging.  We model `species barriers' to prion infection and assess
a related treatment protocol.
\end{abstract}

]

\narrowtext

Transmissible neurodegenerative prion diseases,
such as Mad-Cow Disease (BSE) and Creutzfeldt-Jakob Disease (CJD)
in humans, increasingly represent a serious public health threat\cite{newsci}.
Prusiner and collaborators have shown that the infectious agent (prion)
in these diseases consists of a quantity of a
misfolded form (PrP$^{Sc}$) of the $\sim$200 amino acid PrP$^C$
protein which is
expressed ubiquitously in mammalian neurons\cite{Prusiner1}.
The PrP$^C$ proteins normally reside on
neuron surfaces\cite{PrusinerBio}, and the more hydrophobic PrP$^{Sc}$
forms tend to aggregate: large (micron scale) PrP$^{Sc}$
amyloid plaques are a common post-mortem
feature of brain tissues, and fibrils are observed in {\it in vitro}
experiments.  
Nucleic acid free propagation demands that PrP$^{Sc}$ autocatalyze their
own formation by helping to convert more PrP$^c$ into PrP$^{Sc}$\cite{Prusiner1,Griffith}.
Given that mammalian PrP$^c$ differ by only
5-10\% in amino acid composition and that variant
CJD correlated with BSE has been observed
in England, the efficacy of ``species barriers''
in limiting transmission is of considerable
interest. 

Several facts suggest that prion disease may be driven less by complex biology (as in Alzheimers' disease) than by
physico-chemical processes, including:
(1) The ``universal character'' of sporadic CJD, observed globally
without evident spatio-temporal clustering at an annual background
death rate of 0.5-1.5/million and mean onset age of about 63 years\cite{prionepidem1}. 
(2) Highly reproducible incubation time vs. infectious
dose distributions observed in laboratory animal studies, with a power law relation between mean incubation
time and dose concentration $D$\cite{Prusinerassay}.
(3) The close temporal proximity of disease onset and death.
(4) PrP$^{Sc}$ may be grown {\it in vitro} without biological
processing and is toxic to neurons, though not yet demonstrably
infectious\cite{invitinfect}.
The following question is thus raised: {\it is the incubation time of prion disease
dominated by a fundamental physico-chemical nucleation and growth of PrP$^{Sc}$ protein aggregates?}

We present simulation results for a simple two-dimensional, lattice based
statistical mechanical model for prion disease based at the protein level.
It is all but hopeless to develop
an atomic level simulation capable of spanning the 21 temporal orders
of magnitude between picosecond scale intra-protein
motion and the potentially decades long incubation times of
large prion aggregates.  Our simulation is a bridge between the short distance and time scales
covered in individual protein models and the long time, macroscopic realm of chemical kinetics.
We use our simulations to identify the protectorate of principles necessary to
describe the aforementioned universal features of prion disease.  Specifically, we
find: (1) Concommitant PrP$^{Sc}$ autocatalysis and stable
aggregation can explain the
long disease incubation times, and the difference between sporadic (unseeded) and infectious (seeded) onset.
This principle is consistent with data for yeast prions\cite{yeastdata}, is closely related to the earlier
nucleation theory of Lansbury\cite{lansbury},
and avoids the need to fine tune model parameters with separated autocatalysis and aggregation\cite{eigen}.
(2) The distributions of disease
incubation times (well defined onset edge, broad fluctuation driven
spectrum for dilute prion doses, narrow
short time shifted distribution for more concentrated prion doses) stongly suggest
a central role for growth from a minimally
nanometer scale seed (with order 10 PrP$^{Sc}$ proteins).
With these assumptions we obtain a
one parameter fit to the BSE incubation time data, an order of magnitude
correct estimate of the mean incubation time for BSE and infectious CJD,
and we argue that self-averaging
explains a narrowing of incubation time distributions
observed in the laboratory\cite{Prusinerassay,Lasmezas}.
Finally, by considering both sticking
and interconversion rate differences for prions of different species,
we provide a theoretical underpinning to data for the
species barrier efficacy and to proposed treatments based upon coating
of incipient prion plaques.

Our assumption of two-dimensionality is consistent with the observation
that PrP$^c$ resides on neuron surfaces and autocatalysis and aggregation
must certainly initiate there.  The lattice structure should be
irrelevant at long times. By assuming a triangular lattice we
minimize the anisotropy (which appears first in rank six tensors for hexagonal
lattices).  This simulation yields the crucial qualitative features necessary to specify
the ``prion protectorate.''
Our resulting prion aggregates are essentially amorphous, rather than
quasi-one-dimensional filaments as found in some {\it in vitro}
experiments.  We shall discuss an anisotropic simulation yielding fibrils
elsewhere.

Our model is based upon the energy landscape schematized in
Fig. 1.  Individual lattice cells
take on three values: unoccupied (implicitly filled with water),
occupied by a PrP$^c$ protein, or occupied by a PrP$^{Sc}$ protein.
At the beginning of each simulation, we
choose a monomer PrP$^c$ configuration distributed randomly, with
a number (0-4) of PrP$^{Sc}$ seeds located randomly as well.

In the spirit of cellular and molecular automata, we eschew an explicit Hamiltonian
based molecular dynamics in favor of the following updating rules, specified in time units
of one full lattice sweep ($d$ is the nearest neighbor coordination number):\\

\indent{{\it (1) Identify each occupied cell as PrP$^C$ or PrP$^{Sc}$ by}}\\
\indent{{\it checking its number of nearest neighbors.}}\\
\indent{{\it (2) For each PrP$^C$, move one step in a random di-}}\\
\indent{{\it rection if that adjoining site is unoccupied.}}\\
\indent{{\it (3) Identify aggregates (size $N_P$) by PrP$^{Sc}$ presence;}\\
\indent{{\it move them per Rule (2) with probability $1/\sqrt{N_p}$.}}\\
\indent{{\it (4) For $d=0,1$ the protein remains PrP$^C$. }}\\
\indent{{\it (5) For $d=2$, the protein is
PrP$^C$ or PrP$^{Sc}$ with equal}}\\
\indent{{\it probability}}\\
\indent{{\it (6) For $d\ge 3$ the protein is PrP$^{Sc}$.}}\\

Rules (1-3) ensure diffusive motion of PrP$^C$ monomers and PrP$^{Sc}$ aggregates.
Rules (4-6) specify protein interactions and autocatalysis, and reflect
the relative hydrophobicity of PrP$^{Sc}$.
To assure the biologically plausible condition of constant
average PrP$^C$ concentration, at the start of a new sweep
we randomly place a new PrP$^{C}$ for each lost in the previous update.

This procedure is repeated for up to millions of full sweeps
through the lattice. Note that a sweep
defines the minimum time scale for conversion of PrP$^C$ to PrP$^{Sc}$,
which is presumably of order a second or less in real time.  We
work with maximum lattice sizes of $N=4\times 10^4$.  Practically, we are able to
run at areal monomer concentrations $c\ge$ 0.1\%, which are likely to be about three
orders of magnitude larger than in the brain.

We run our seeded simulations repeatedly following the largest nano-aggregates until they
reach a size of up to 0.4\% of the lattice.  Each seed consists of 10 PrP$^{Sc}$ monomers\cite{minimalseed}
For a given aggregate size, we have studied
the properties of the distribution of lifetimes to reach that size.  For a final
aggregate size of ${\cal A}$=0.2\%$N$ (80 monomers) we display the corresponding distributions $P_D$
in Fig. 2(a)
for several values of $c$ and $D=1$.
$P_D$ displays greater
fluctuations for small $c$, and increasing
$c$ shifts $P_D$ to short times.
Fluctuations dominate the growth from the small seed aggregate.
For larger ${\cal A}$ we find that the onset time grows
sublinearly in ${\cal A}$, while the distribution width grows weakly with ${\cal A}$.
To within fluctuations, the $P_D$ curves for different $c$ collapse when scaled by
the mean time $t_m$. (The curves broaden somewhat for $D=2,3,4$.)
This implies the scaling law
\begin{equation}
P_D(t,c) \approx {1\over t_m(c)} F_D({t\over t_m(c)})
\end{equation}
In the range of
concentrations studied, we can fit $t_m(c)$ well by either: 
(a) the polynomial
form $t_m(c)\sim Ac^{-1} + Bc^{-2}$ (the first term reflecting
monomer aggregation, the second representing dimer aggregation), or 
(b) a
power law form, $t_m(c)\sim c^{-\alpha(D)}$, with 
$\alpha(D) = 1.66(D=1),~1.49(D=2)$. 
For finite ${\cal A}$ and $c\to 0$ the
exponent $\alpha(D)$ should tend to two, reflecting dominant dimer aggregation.
Fit (b) is consistent with a nontrivial scaling limit as 
${\cal A}\to \infty$, $c\to 0$. 
Both fits to $t_m(c)$ give order of magnitude equivalent
extrapolations to lower $c$ relevant to biological contexts. 
Assuming $t_m(c) \sim D^{-\beta}$ at fixed $c$, we estimate $\beta\le 1$ with 
large numerical uncertainty. 
Taken together, these data lead us to a conjecture of
{\it asymptotic compression} of $P_D$:  for $c\to 0$, an increasingly large
range of ${\cal A}$ will have essentially the same scaled shape, because the time $t_2$
it takes
to go from ${\cal A}$ to say,  $2{\cal A}$ will be much smaller than $t_m$
for going from $10\to {\cal A}$.

We believe that other processes, such as protease attack or competition between PrP$^{Sc}$-PrP$^{Sc}$
and PrP$^{Sc}$-neuron binding will limit prion aggregation on the neural surface and cause
fissioning and spreading of nano-aggregates to other neurons.  Such fission
is necessary for exponential growth of infectious agents, as noted
previously\cite{polymer}.
At the order-of-magnitude level $t_2$ sets the doubling time for disease
spread, assuming that inter-cellular diffusion rates exceed intra-cellular ones.
Given $t_2<<t_m$ (at biological $c$ values) both in our fission hypothesis and experiment, slow growth
fluctuations for small seed aggregates can dominate the overall incubation time for disease.
For example, in hamster experiments the first observable levels of PrP$^{Sc}$
in brain tissues occur at about 90 days, with symptom onset at 120 days, and a doubling time of 2 days\cite{hamsterdoubling}.
Given the conjectured asymptotic compression, the incubation time distribution will be relatively
insensitive to a wide range of nano-aggregate sizes for fission parents or progeny.

We stress several points of agreement with observation and laboratory
data.  First, we always find
death (runaway growth of a killer aggregate) in the presence of infection,
in agreement with experiment and the aggregate nucleation hypothesis
for prion disease. Second, our rough scaling of $t_m$ with $D$
agrees qualitatively with the extraordinarily reproducible
dose vs. incubation time data of Prusiner and collaborators for prion
infection in laboratory mice, for which $\tau_I \sim D^{-.62}$.
Third, our lifetimes are in order of magnitude agreement with
observed BSE incubation times and new variant CJD death ages,
which we can prove by exploiting our concentration
scaling.  For a single seed, at our lowest
concentration of 0.1\%  our distribution peaks at
about 10$^4$ sweeps.
Assuming a literature estimate of
concentration of {\it in vivo} prions in solution
of 100 nanomole/liter\cite{laurent} and a protein diameter of order 
one nanometer 
suggests a dimensionless areal concentration of about
$10^{-3}$\%.  With the above concentration scaling
this gives an incubation time of about
10$^{8}$ sweeps.  If we take the 5 year mean incubation time for
BSE this implies our basic misfolding time (one sweep) is about
one second, which is reasonable at the order of magnitude level.
Finally, by simply scaling
time for ${\cal A}=80$  by the mean time $t_m$, we achieve an exceptionally
good fit (${\cal A}$) of
the inferred incubation  time distribution
for BSE in the United Kingdom (for cattle born in 1987)\cite{bseepidem} to our
scaling curve as shown
in Fig. 2(b).  This supports
our hypothesis that runaway autocatalytic aggregation seeded by a few prion
nano-aggregates triggers the disease. This is very plausible since
(i) infection is believed to have come from rendering plant offal
derived from many animals after considerable processing,leading to
substantial dilution, and (ii) prion transmission by oral consumption 
is estimated to be less efficient than innoculation by a factor of
10$^9$.

Sporadic CJD develops in the absence of any infection or genetic
predisposition to prion disease.  We have studied the lifetime
distributions in
this case and find that they are very different in shape than the
seeded runs, being much broader and dimer fluctuation dominated (peaks
scale as $c^{-2}$). For our estimated biological concentration of
$c=10^{-3}$\%, this could give a mean onset time as large as 10$^3$ years.
Assuming a constant height tail from the minimum single seed onset to
the mean human lifespan taken with the 1 part in 10$^6$ odds of dying
from sporadic CJD gives an estimate of a total death probability of
a part in 10$^5$ up to this mean on-set time.  This is consistent with
our simulations, but not resolvable currently.

In contrast to the breadth of our single seed lifetime distributions, in the
event of dosing by a greater number of seeds, self-averaging will narrow the distribution,
collapsing it towards the onset (shorter times).  Assuming growth
proceeds independently for different seeds (which is quite reasonable
for the low concentrations in living organisms) distributions would
narrow by $\sqrt{D}$.   Several {\it in vivo} experiments show distribution time narrowing with
increased doseage, but are not amenable to quantitative analysis at this time\cite{Prusinerassay,Lasmezas}.

We turn now to the species barrier. To simulate interspecies transmission
we replace the integer-valued coordination, the key parameter which determined
the stability of different protein conformations in the earlier studies, by
a new variable $x=N+N^{\prime}\times P$ for PrP$^C$
and $x^\prime=N^\prime+N\times P^\prime$, for PrP$^{C\prime}$,
where $N$ is the number of PrP$^{C,Sc}$ neighbors and $N^{\prime}$ the number of
PrP$^{C,Sc\prime}$ neighbors. The parameters $P$ and $P^\prime$ (chosen less than unity)
provide a measure of the reduced
effectiveness of interspecies conversion.
To deal with the continuous variable $x$, rather than the integer coordination used earlier, we
represent
the PrP$^C$ to PrP$^{Sc}$ conversion probability $f(x)$ by
\begin{equation}
f(x) ={ {1} \over {e^{\beta (x-2) } + 1}}
\end{equation}
with a similar definition for PrP$^{C\prime}\to$PrP$^{Sc\prime}$
conversion.  A separate parameter describes physisorption
to an aggregate of alien prions.  We allow a
PrP$^{C,Sc}$ cluster of mass $M$ ($M=1$ being a monomer) to
desorb from an alien prion aggregate with probability $q^K / \sqrt{M}$,
where $q$ is the physisorption parameter, and $K$ is the number of units in the
cluster adjacent to the alien aggregate.  We
define $q^{\prime}$ for PrP$^{C,Sc\prime}$ similarly.
Thus, we describe species barriers in terms of the two parameters
affecting conversion rates ($P,P^{\prime}$) and physisorption ($q,q^{\prime}$).
Making $P\ne P^{\prime}$ and $q\ne q^{\prime}$ allows for an asymmetry
in infectivity, and we can model one of the most striking aspects of the species barrier,
namely its asymmetry ({\it e.g.} mice infect hamsters well, hamsters infect mice poorly).

We have carried out several simulations relevant to the species barrier.
First, we have placed alien prion `walls' on a boundary,
representing a single large aggregate from another species.
With suitable choice of parameters, this leads to the asymmetric species barrier
between two species.
Namely, disease onset
in one species (with large aggregates) and only sub-critical PrP$^{Sc}$
concentration near the wall in the other. Recent experimental data has also shown the build up
of such sub-critical PrP$^{Sc}$ concentration in the asymmetric interspecies
transmission \cite{subclinical}.

Second, suppose the asymmetry is such that $P<<P^{\prime}$.  In this
case, PrP$^{Sc\prime}$ is more favorably formed from PrP$^{C\prime}$
in the presence of a seed aggregate of PrP$^{Sc}$ than PrP$^{Sc}$ is
formed from PrP$^{C}$ in the presence of a PrP$^{Sc\prime}$ seed.
Thus, by injecting a large enough initial dose of alien PrP$^{C\prime}$
proteins into the organism, it should be possible for PrP$^{C\prime}$ to compete favorably
with PrP$^{C}$ for aggregation, and thus extend the incubation time.
Such protocols have been tested experimentally: (i)
Experiments with the coating dye molecule Congo Red\cite{congored}
reveal a surprising {\it non-monotonic} dose vs. lifetime
relation:  small Congo Red concentrations yield a reduced time
for incubation, while larger Congo red concentrations significantly
extend the incubation time.  (ii) {\it In
vitro} experiments show that the above coating scenario with alien
prions works\cite{aliencoat} when the initial PrP$^{C\prime}$ concentration
$c^{\prime}> c$.

We find both of these experimental features in our simulation
as shown in Fig. 3.  For $c'/c < 2$, the incubation time is
slightly shortened,
while for $c'/c > 2$ the incubation time is increased.  We find
that the lifetime shortening arises because once a few PrP$^{C\prime}$
aggregate, they partially block the motion of adjacent
PrP$^{C}$ proteins, enhancing the misfolding probability of the latter.
However, further PrP$^{C}$ misfolding is significantly reduced once PrP$^{Sc\prime}$ coats
the seed.

A crucial test of our theory would be to monitor, {\it in vitro},
the lifetime distributions of normal prion proteins seeded by small
aggregates as a function of both $c$ and $D$ to look for the scaling
behavior. If it is not possible
to easily achieve $D=1,2$, presumably the distribution of Fig. 2(b)
could be estimated by extrapolation from successively diluted
PrP$^{Sc}$ doses.  This would also test the extent to which the
self-averaging applies.   If our single seed hypothesis is correct,
it is important to assess how passage occurs upon infection.

Theoretically, it is important to extend our model by including internal
degrees of freedom for the proteins which can provide an orientational
dependence to the aggregation, allowing for linear fibril formation as
opposed to amorphous aggregates.  This connects naturally with 
simulations on individual proteins, and
also allows for  
the study of different protectorates for collective protein phenomena.

{\it Acknowledgements}.  We acknowledge useful discussions with
S. Carter, F. DeArmond, R. Fairclough, C. Ionescu-Zanetti, T. Jue, R. Khurana, and
C. Lasm\'{e}zas.  F. P\'{a}zm\'{a}ndi has been supported by National Science
Foundation Grant DMR-9985978.  R.R.P.S. and D.L.C. have benefitted from
discussions at workshops of the Institute for Complex Adaptive
Matter.  We are grateful for a grant of supercomputer time from the Lawrence
Livermore National Laboratory.

\begin{figure}
\epsfxsize=2.5in
\epsffile{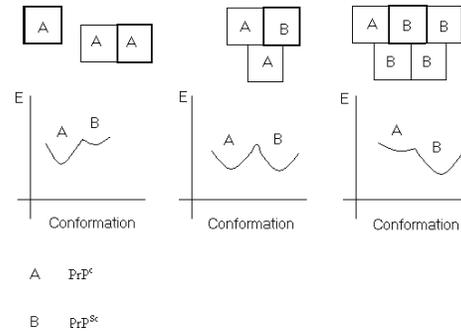}
\vspace*{0.05cm}
\caption{ Schematic energy landscapes for 
model prion proteins.  Upper figures show proteins in varying coordination
environments corresponding to rules (1-6), 
lower figures show schematic energy landscapes.  }
\end{figure}

\begin{center}
\begin{figure}
\psfig{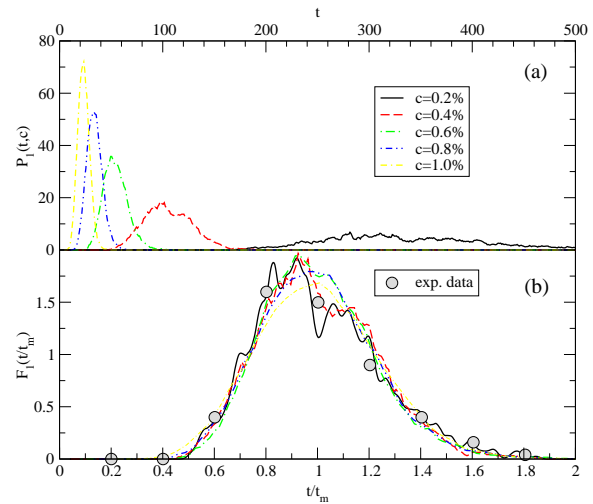}
\vspace*{0.05cm}
\caption{ Aggregation lifetime distributions $P_D(t,c)$ for a single seed ($D$=1).
Seeds have ten proteins, simulations terminate at aggregate size ${\cal A}=80$,
time is measured in units of 10$^4$ full lattice sweeps.   (a)
Dependence upon areal concentration $c$ of normal prion proteins.
(b) Scaled distribution $F_1(t/t_m)$ vs. $t/t_m$.  $t_m$ 
is the mean lifetime for a given concentration.
Points are inferred incubation time data for BSE-infected
cattle in the United Kingdom born in 1987.    }
\end{figure}
\end{center}

\begin{center}
\begin{figure}
\psfig{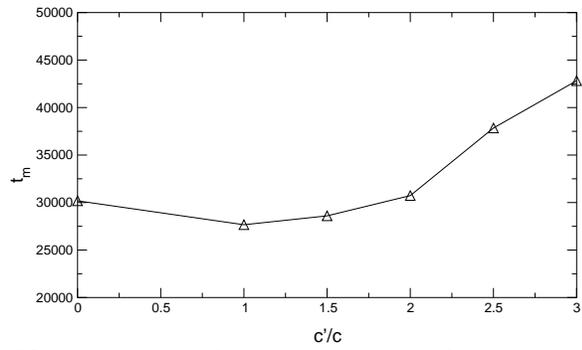}
\vspace*{0.05cm}
\caption{  Mean incubation time $t_m$ vs.
relative concentration of alien prions $c\prime/c$ (test of coating
treatment; see text for discussion).  }
\end{figure}
\end{center}


\begin{references}
\bibitem{minimalseed} The
minimally stable seed in our model has seven monomers, but exhibits spurious barriers for
transitioning from seven to ten monomers.
\end{references}
\end{document}